\documentclass[aps,pre,twocolumn]{revtex4}
\usepackage{graphicx}
\usepackage{amsmath}
\usepackage{braket}
\usepackage{hyperref}
\usepackage{subfigure}
\usepackage[usenames,dvipsnames]{color}

\newcommand{\ct}{\cite}
\newcommand{\la}{\lambda}
\newcommand{\bi}{\bibitem}
\newcommand{\be}{\begin{equation}}
\newcommand{\ee}{\end{equation}}
\newcommand{\ba}{\begin{eqnarray}}
\newcommand{\ea}{\end{eqnarray}}

\begin{document}
\title{Phase Transition in the periodically pulsed Dicke Model}
\author{Sayak Dasgupta}
\affiliation{Department of Physics, Indian Institute of Technology, 208016, Kanpur}
\author{Utso Bhattacharya}
\affiliation{Department of Physics, Indian Institute of Technology, 208016, Kanpur}
\author{Amit Dutta}
\affiliation{Department of Physics, Indian Institute of Technology, 208016, Kanpur}
\begin{abstract}
We study the effect of pulsed driving and kicked driving of the interaction term on the non-equilibrium phase transition in the Dicke Model. Within the framework of Floquet theory, we observe the emergence of new non-trivial phases on impingement by such periodic pulses. Notably, our study reveals that  a greater control over the dynamical quantum criticality is possible through the variation of multiple parameters related to the pulse, as opposed to a single parameter control in a monochromatic drive. Furthermore, the probability
of the system remaining trapped in a metastable state during the observed first order transition from the super-radiant to normal phase is found to be higher for small number of kicks (or pulses) in comparison to the sinusoidal perturbation.
 \end{abstract}
\maketitle
\section{Introduction}
\label{sec_intro}
The study of quantum phase transitions (QPT) \cite{sachdev99} is well documented in cases where a system is slowly driven through a quantum critical point (QCP).  A QCP is characterized by a  diverging relaxation time and correlation length and hence the changes due to such non-adiabatic crossing is reflected, most strongly, in defects that are generated  in the final state. This is described by the Kibble-Zurek mechanism (for review see, \ct{polkovnikov11,dutta10,dziarmaga11}). 

At the same time, there have been numerous studies on the periodic dynamics of closed quantum systems \ct{mukherjee09,das10,russomanno12}. The motivation behind these studies arise from
the possibility of experimentally realizing a topologically non-trivial phase by application of light on a topologically trivial phase; in this context the most important materials to have caught attention are Floquet graphene \ct{oka09,kitagawa10} and the Floquet topological insulator \ct{linder11,rechtsman13} (for a review see \ct{dora13}). Periodically driven systems have also been explored from the view point of dynamical steady state \ct{russomanno12,sharma14}, dynamical freezing \ct{das10} and dynamical localization \ct{lazarides14,nag14}.

The question that arises is whether one can talk of non-equilibrium phase transitions in a periodically driven system, like the periodically pulsed Dicke model (DM). The DM \ct{dicke54,
emary03_prl,emary03,vidal06} is a system  of ``$N$" interacting 2-level atoms placed in a bosonic cavity with a coupling characterized by the parameter $\lambda_{o}$. It is widely used in quantum optics to study collective effects. The DM in the thermodynamic limit, in the absence of driving, shows a second-order quantum phase transition from a normal phase (NP) which is microscopically excited to a super-radiant (SR) phase which is macroscopically excited beyond a critical value of the coupling between the spins and a single frequency Bosonic mode $\lambda_{o} (= \lambda_{c})$. This transition has been characterized using quantum information theoretic measures \ct{zanardi06,bhattacharya14} as well as information geometry \ct{tapo}. In the case of a finite sized DM the SR phase shows a chaotic behaviour which was characterized using quantum information theoretic measures in \ct{bhattacharya14}. As the system is taken to the thermodynamic limit the chaotic behaviour of the super radiant (SR) phase gives way to a well-defined QPT. 

Significant progress was made by Bastidas $et~ al.$\cite{bastidas12} in characterizing such non-equilibrium behaviour in the Dicke Model(DM), where they introduced a harmonic time varying coupling constant $\lambda = \lambda_{o} + \lambda_{1}\cos(\Omega t)$. This external drive affected the parameters of the static system in a manner, such that only a slight shift in the critical point was observed. The primary feature of such an external driving is that transitions to excited states become rampant, thus the phases depend on the manner in which the non-equilibrium conditions are employed. Only a slight shift in the critical value, in this non-equilibrium scenario is indicative of the fact that in spite of driving, the system still remains close to equilibrium, ensuring well defined quantum phases. They used the resulting lowest quasi energy (LQE) landscape, obtained from the Floquet Hamiltonian of the driven DM to characterize the different dynamically generated phases. The shift in the QCP, occurrence of side-band QPTs, and the appearance of new novel phases are some of the consequences of the periodic driving. 

The monochromatic drive as discussed in \cite{bastidas12} can indeed be used to effectively generate the novel dynamical phases. However, it has the disadvantage of allowing control through a single parameter only. To address this issue, in this work, we focus on a  Dicke Hamiltonian (DH) with a pulsed square wave interaction term and investigate its non-equilibrium transition scenario, providing the necessary comparison with the monochromatic case. Though non mono-chromatic, which makes a mathematical treatment more complicated, the square wave pulses allow greater control in generating the non trivial phases mentioned above. In our present work we study the Dicke Hamiltonian (DH) with a pulsed square wave interaction term. We present a detailed study of the extreme asymmetric limit of the square wave pulse, the Dirac-Delta comb, in order to highlight the techniques involved, following up with a more general treatment of a square wave pulse which is asymmetric in time, showing that this case indeed reduces to the former in the correct limits.
 
 The remaining article is organised as follows; we provide a brief review of the DM in section \ref{sec_model} while in section \ref{sec_Floq}, a brief discussion of Floquet theory is provided concentrating on the Floquet operator for piece-wise (in time) continuous perturbations. In section \ref{sec_KDM}, the emergence of instabilities due to kicking in the model in the NP of the DH is studied. In section \ref{sec_rot_frame}, we work in the rotating frame, and derive the LQE surface from the transformed Hamiltonian, following it up by calculating and analyzing the LQE surface for an asymmetric square wave pulse drive (`Bang-Bang') in section \ref{sec_BB}. Finally we present our concluding remarks in section \ref{sec_conclusion}. \\
\section{The Dicke Model: A Review}
\label{sec_model}
We present here a brief review of the Dicke Hamiltonian (DH) which describes a single mode bosonic field interacting with an ensemble of $N$  two level atoms \ct{dicke54}, given by:
\begin{equation}
H = \omega_{0}\sum_{i=1}^{N}s_{z}^{i}+\omega a^{\dagger}a+\sum_{i=1}^{N}\frac{\lambda_{o}}{\sqrt{N}}(a^{\dagger}+a)(s_{+}^{(i)}+s_{-}^{(i)})\hspace{0.3cm}[\hbar=1].
\label{eq.DH}
\end{equation}
Here $\omega_{0}$ is the energy level splitting between the two-level systems. $a^{\dagger}(a)$ is the creation (annihilation) operator for the bosonic field; with $[a^{\dagger},a]=1$. In our case, we consider only a single bosonic mode which interacts with $N$ two-level atoms with the interaction strength $\lambda_{o}$.The $i$-{th} atom is described by the spin-half operators $\left(s_{k}^{i};k=z,\pm\right)$, obeying the commutation rules $[s_{z},s_{\pm}]=\pm s_{\pm}$; and $[s_{+},s_{-}]=2s_{z}$.
The origin of the factor $1/\sqrt{N}$ in the interaction term results from the dipole interaction which is proportional to $1/\sqrt{V}$, where $V$ is the volume of the cavity. Taking into consideration that the density of atoms in the cavity is $\rho =N/V$, we find that the coupling strength is of the form  $\lambda/\sqrt{N}$. The scaling factor $\sqrt{N}$ appearing in the interaction plays an important role for the finite ``size" system.

The DH (Eq.~(\ref{eq.DH})) is further simplified by using collective atomic operators,
\begin{equation}
J_{z}\equiv\sum_{i=1}^{N}s_{z}^{(i)};\hspace{4mm}J_{\pm}\equiv\sum_{i=1}^{N}s_{\pm}^{(i)},
\label{eq.ang_mom}
\end{equation}
which obey the usual angular momentum commutation relations.
Here, $j$ is assigned its maximum value $j=N/2$, and this value is constant for a fixed value of $N$. Thus, the $N$ two-level system effectively gets reduced to  a $(2j+1)(=(N+1))$ level system.
The final form of the single-mode DH then looks like,
\begin{equation}
H = \omega_{0}J_{z}+\omega a^{\dagger}a+\frac{\lambda_{o}}{\sqrt{2j}}(a^{\dagger}+a)(J_{+}+J_{-}).
\label{eq.DH2}
\end{equation} 
The resonance condition, $\omega=\omega_{o}$, has been used in the rest of the paper.
The parity operator $(\Pi)$ can be defined here in terms of the total number of excitation quanta $(\hat{N})$ in the system, as
\begin{equation}
\Pi=\exp{\{i\pi\hat{N}\}};\hspace{1cm}\hat{N}=a^{\dagger}a+J_{z}+j,
\label{eq.parity_op}
\end{equation}
Clearly, the operator $\Pi$ can have only two eigenvalues $(\pm 1)$, $N$ being even or odd.  Thus, the DH turns out to be parity conserving as $[H,\Pi]=0$ and, correspondingly the Hilbert-space of the total system is split into two non-interacting sub-spaces. The ground-state has an even parity as it has no finite excitations and $J_{z} = -j$. 

The DH shows a QPT in the thermodynamic limit (as $N$ $\rightarrow \infty$) at a critical value of the atom-field coupling strength $(\lambda_{o})$, $\lambda_{c}=\sqrt{\omega\omega_{o}}/2$ where the symmetry associated with the parity operator $(\Pi)$ is broken. The second derivative of the ground state energy per $j$ with respect to $\lambda_{o}$ shows a sharp discontinuity at the point $\lambda_{o} = \lambda_{c}$ clearly marking the occurrence of a phase transition; this transition separates the NP (for $\la_{o} <\la_c$) from the SR phase (for $\la_{o} > \la_c$). The system in the NP is only microscopically excited whereas the SR phase shows macroscopic excitations.\\
To exactly diagonalise the Hamiltonian in the thermodynamic limit one resorts to the Holstein-Primakoff transformation (applied to the DH as in \cite{lambert04}) of the angular momentum operators, given by :
\begin{eqnarray}
J_{+} &=& b^{\dagger}\sqrt{2j-b^{\dagger}b};\\ \nonumber
 J_{-} &=& \sqrt{2j-b^{\dagger}b}\hspace{0.1cm}b;\\ \nonumber
J_{z} &=& \left(b^{\dagger}b-j\right);
\label{eq.Holstein_Primakoff}
\end{eqnarray}
where $[b,b^{\dagger}]=1$. With these substitutions we get the DH as:
\begin{eqnarray}
H &=& \omega_{0}\left(b^{\dagger}b-j\right) + \omega a^{\dagger}a \\ \nonumber
&+& \lambda\left(a^{\dagger} + a\right)\left(b^{\dagger}\sqrt{1-\frac{b^{\dagger}b}{2j}}+\sqrt{1-\frac{b^{\dagger}b}{2j}b}\right).
\label{eq.HP_DH}
\end{eqnarray}
In the thermodynamic limit $j\to\infty$ in the NP, the expression reduces to:
\begin{eqnarray}
H &=& \omega_{0}\left(b^{\dagger}b-j\right) + \omega a^{\dagger}a \\ \nonumber
&+& \lambda\left(a^{\dagger} + a\right)\left(b^{\dagger}+b\right).
\label{eq.HP_DH_ab}
\end{eqnarray}
In the SR phase to capture the macroscopic occupations of both the field and the atomic ensembles we have to displace the bosonic modes in Holstein-Primakoff representation, in either of the following ways,
\begin{eqnarray}
a^{\dagger}\rightarrow c^{\dagger} + \sqrt{\alpha};\hspace*{1cm}b^{\dagger}\rightarrow d^{\dagger} - \sqrt{\beta};\\ \nonumber
a^{\dagger}\rightarrow c^{\dagger} - \sqrt{\alpha};\hspace*{1cm}b^{\dagger}\rightarrow d^{\dagger} + \sqrt{\beta};
\label{eq.displacement}
\end{eqnarray}
with $\sqrt{\alpha} = X\sqrt{j}$ and $\sqrt{\beta} = Y\sqrt{j}$, and retaining only the terms linear in $j$. Both the choices of the bosonic displacements give identical Hamiltonians. Hence, every state is doubly degenerate in the SR phase.\\

\section{Floquet Theory:}
\label{sec_Floq}
The Floquet technique (a temporal version of Bloch's theorem) \cite{shirley65,griffoni88,stockmann99} is meant to deal with Hamiltonians subjected to a time-periodic potential of the form $H = H_{0} + V(t)$, where $V(t+\tau)=V(t)$. A discrete time translation operator $T$ can be introduced, such that $T\psi_{n}(x,t)=\psi_{n}(x,t+\tau)=\lambda_{n}\psi_{n}(x,t)$. For the solution to be stationary $\lambda$ has to be a pure phase of the form $e^{-i\phi_{n}}$. Thus we have a solution of the form :
\begin{equation}
\psi_{n}(x,t+\tau) = e^{-i\omega_{n}t}u_{n}(x,t),
\label{eq.floq1}
\end{equation}
where $u_{n}(x,t+\tau)=u_{n}(x,t)$ and $\omega_{n} = \phi_{n}/\tau$. Just as in the case of Bloch's theorem, one obtains quasi momenta $\vec{k}$ here, one obtains quasi energies of the form $E_{n} = \hbar \omega_{n}$, defined within the first Brillouin zone, $\langle-\frac{\hbar}{2\tau},\frac{\hbar}{2\tau}\rangle$.  It is advantageous to observe the system after intervals of the time period $\tau$ when the unitary evolution operator is given by, $U(n\tau,0)=[U(\tau,0)]^{n}$. We can directly find the Floquet quasi-energies from the diagonal representation of $U$, given by $U_{D} = diagonal[e^{-i\phi_{n}}]$. 

It is to be noted that for a generic periodic perturbation finding the eigenphases of $U$ is cumbersome as the Fourier transform of the potential usually contains an infinite number of modes giving rise to an infinite matrix which can be diagonalised only under restrictions like rotating wave approximation (RWA). The situation however becomes tractable when we have periodic $\delta$-function
kicks $V(t) = V_{0}\sum_n \delta(t-n\tau)$. Here the Hamiltonian can be made piece-wise integrable by using the potential:
\begin{eqnarray}
V(t) &=& 0 , \hspace*{1.5cm} 0<t<\tau-\Delta\tau \\ \nonumber
V(t) &=& V_{0}/(\Delta\tau), \hspace*{.4cm} \tau-\Delta\tau \leq t < \tau.
\end{eqnarray}
Thus on integrating and taking the limit $\Delta\tau \rightarrow 0$ we find the exact form $U(\tau,0)= \exp[-i\int_{0}^{\tau}H(t)dt]=\exp(-i V_{0})\exp(-i H_{0}/\tau)$.

\section{Kicked Dicke Model}
\label{sec_KDM}
We first present the case of the kicked (Dirac comb in time) DM as it provides a mathematically easier platform which can be used to highlight the techniques involved in studying the DM in the presence of a non monochromatic interaction term.\\
In the static DM, as discussed in the previous section, the atomic ensemble interacts with the bosonic mode of frequency $\omega$, through a time-independent dipole interaction of strength $\la_{o}$. We modify this in the current section by making the interaction strength time dependent. Thus the DH as shown in Eq.~(\ref{eq.DH2})  now includes a kicked interaction term:
\begin{equation}
\label{eq.KDH}
H = \omega_{0}J_{z}+\omega a^{\dagger}a+\frac{\lambda(t)}{\sqrt{2j}}(a^{\dagger}+a)(J_{+}+J_{-})
\end{equation}
with $\lambda(t) = \lambda_{o} + \lambda_{1}\sum_{k=1}^{\infty}\delta(\frac{t}{T}-2\pi k)$. It should be noted here that the Hamiltonian  retains its parity symmetry even in the presence of $\delta$-kicks. Let us now introduce the position and momentum operators for the two bosonic modes $(a,b)$:
\begin{eqnarray}
\label{eq.xy}
x &=& \frac{1}{\sqrt{2\omega}}(a^{\dagger} + a); \hspace*{3mm} p_{x} = i\sqrt{\frac{\omega}{2}}(a^{\dagger} - a).\\ \nonumber
y &=& \frac{1}{\sqrt{2\omega}}(b^{\dagger} + b); \hspace*{3mm} p_{y} = i\sqrt{\frac{\omega}{2}}(b^{\dagger} - b).
\end{eqnarray}
where $(x,p_{x})$ are the quadratures of the cavity field and $(y,p_{y})$ those of the atomic ensemble. The Hamiltonian can now be written in terms of these quadratures as:
\begin{equation}
\label{eq.DH_XY}
H(t) = \frac{1}{2}\left(p_{x}^{2} + \omega^{2}x^{2} + p_{y}^{2} + \omega^{2}y^{2}\right) +2\omega xy -\omega\frac{(N+2)}{2}.  
\end{equation}
From this we can obtain the Heisenberg equation of motion for the quadratures $q_{\pm} = (x(t) \pm y(t))/\sqrt{2}.$ as:
\begin{equation}
\label{eq.motion}
\ddot{q}_{\pm}(t) = -\left(\epsilon_{\pm}^{2} \pm 2\omega\lambda_{1}\sum_{k=1}^{\infty}\delta\left(\frac{t}{T}-2\pi k\right)\right)q_{\pm}(t)
\end{equation}
with $\epsilon_{\pm} = \sqrt{\omega^{2} \pm 2\lambda_{o}\omega}$, being the excitation energies in the NP of the undriven Hamiltonian. Unlike the case in \cite{bastidas12}, where  an evolution equation of the form of a Mathieu equation 
was obtained and  the stability analysis of the solutions was carried out using standard procedures, our evolution equation Eq.(\ref{eq.motion}), contains a delta function in time which can only be handled through a discrete map based stability analysis.
 To analyse Eq.~(\ref{eq.motion}) we resort to the following scheme: for each mode $q_{\pm}$ we allow evolution of $q_{\pm}(t)$ under the unkicked Hamiltonian (which resembles a simple harmonic oscillator Hamiltonian for each mode in the $(q,p)$ representation), followed by pulse whose width is given by $\Delta t$ (later taking the limit $\Delta t \to 0)$. This comprises one full cycle. Thus a map is created in which we follow the evolution from $q_{n}$ to $q_{n+1}$, each cycle comprising a single kick. The map obtained is:
\begin{eqnarray}
\label{eq.map.minus}
q_{n+1}^{-} = q_{n}^{-}\cos(\epsilon T) +\left(\frac{p_{n}^{-}+2\omega\lambda_{1}q_{n}^{-}}{\epsilon}\right)\sin(\epsilon T). \\ \nonumber
p_{n+1}^{-} = -q_{n}^{-}\sin(\epsilon T) + (p_{n}^{-} + 2\omega\lambda_{1}q_{n}^{-})\cos(\epsilon T)
\end{eqnarray}
\begin{figure}
\includegraphics[height= 5cm, width=6cm]{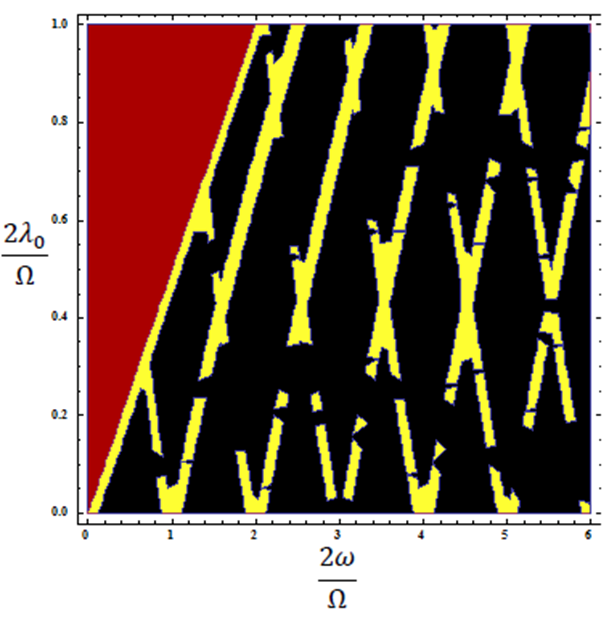}
\caption{(color online) This figure shows the unstable regions in the parameter space for the maps derived from Eq.\ref{eq.map.minus} and its counterpart for the $q_{+}$ mode combined. The black background region represents the NP while the red background represents the SR phase, the yellow striations represent the unstable zones. We expand and perform the energy surface calculations around $\epsilon_{\pm} = m\Omega/2$, for $m=0$. That is the region for obtained at a low $\lambda_{0}$.}
\label{fig_stability}
\end{figure}
Equations connecting $q_{n+1}^{+}$ to $q_{n}^{+}$ can be obtained in a similar fashion. To determine stability we find the eigenvalues $\lambda_{1,2;\pm}$ and use the criterion $|\lambda|>1$ for unstable solutions. When both the normal modes $q_{\pm}$ are stable the NP Hamiltonian allows for bound solutions around microscopically excited atomic and field modes, while the unstable solutions correspond to macroscopic excitations. We see the occurrence of zones of instability in the NP as shown in Fig.\ref{fig_stability}. The first occurrence of these zones satisfies the resonance condition between the kicking frequency and the unkicked excitation energies; $\epsilon_{\pm} = m\Omega/2$ where $\Omega = 2\pi/T$ is the  frequency of kicking. In the zones of stability the $q_{\pm}$ modes effectively describe a system with no microscopic excitations. In the zones where the solution becomes unstable 
evidently  the effective Hamiltonian can not be  described by  Eq.~(\ref{eq.DH}) and 
we must allow for the possibility of macroscopic excitations and arrive at  an effective Hamiltonian describing these excitations around the zones of instability. Thus the de-stabilization of the solutions is indicative of the onset  of `non-equilibrium' phase transitions.

\section{The Rotating Frame:}
\label{sec_rot_frame}
We now produce an analysis of the Hamiltonian (\ref{eq.KDH}) around the unstable zones obtained in the previous section. This enables us to obtain the form of the lowest quasi energy surface in this region from where we can further study the nature of the non-equilibrium QPT. The difficulty in analyzing a strongly coupled system can be reduced by transforming  to a rotating frame, that converts the interaction amplitudes into phases. Hence, in the rotating frame, the rotation is given by the unitary operator:
\begin{equation}
\label{eq.U_rot}
U_{m}(t) = \exp\left[-i\gamma(a^{\dagger} + a)J_{x}\right]\exp\left[-i\theta_{m}(a^{\dagger}a + J_{z})\right], 
\end{equation}
where $\theta_{m} = \frac{m\Omega}{2}t$, and $\gamma = \frac{1}{\sqrt{N}}\frac{\lambda_{1}}{\Omega}k$. Here, we utilize the fact that for a small static coupling $\lambda_{o}$, the $m^{th}$ instability zone arises close to $\omega \approx m\Omega/2$, see Fig.(\ref{fig_stability}). In this frame the Floquet Hamiltonian (given by $H_{f} = H - i\delta_{t}$) separates into modes given by $H_{m}= U_{m}^{\dagger}H_{f}U_{m}$. The explicit form is obtained as:
\begin{eqnarray}
\label{eq.flo_ham_modes}
H_{m}(t) &&= \delta^{m}a^{\dagger}a \\ \nonumber
&&+\left[\omega\cos(\gamma\left(a^{\dagger}e^{i\theta_{m}} + ae^{-i\theta_{m}}\right))-\frac{m\Omega}{2}\right]J_{z} \\ \nonumber
&&+\frac{\la_{o}}{\sqrt{N}}\left(a^{\dagger}e^{i\theta_{m}} + ae^{-i\theta_{m}}\right)\left(J_{+}e^{i\theta_{m}} + J_{-}e^{-i\theta_{m}}\right)\\ \nonumber
&& -i\frac{\omega}{2}\gamma\left(a^{\dagger}e^{i\theta_{m}} - ae^{-i\theta_{m}}\right)\left(J_{+}e^{i\theta_{m}} + J_{-}e^{-i\theta_{m}}\right) \\ \nonumber
&&+\frac{\omega}{4}\gamma^{2}\left(J_{+}e^{i\theta_{m}} + J_{-}e^{-i\theta_{m}}\right)^{2} \\ \nonumber
&&-i\frac{\omega}{2}\sin\left[\gamma\left(a^{\dagger}e^{i\theta_{m}} + ae^{-i\theta_{m}}\right)\right]\left(J_{+}e^{i\theta_{m}} - J_{-}e^{-i\theta_{m}}\right)
\end{eqnarray}
with $\delta^{m} = \omega - \frac{m\Omega}{2}$. Since in our calculations the factor $\gamma$ is time independent we do not have to resort to a rotating wave approximation (RWA) unlike the case in \cite{bastidas12}. The simplest possible thing that can be done to investigate the LQE surface is to expand the Hamiltonian given in Eq.~(\ref{eq.flo_ham_modes}) around the $m = 0$ instability zone such that in our calculation $\lambda_{o}$ remains small. Then we obtain:
\begin{eqnarray}
\label{eq.hoo}
H_{m=0}(t) &&= \omega a^{\dagger}a +\omega\cos(\gamma\left(a^{\dagger} + a\right))J_{z} \\ \nonumber
&&+\frac{\la_{o}}{\sqrt{N}}\left(a^{\dagger} + a\right)\left(J_{+} + J_{-}\right)\\ \nonumber
&& -i\frac{\omega}{2}\gamma\left(a^{\dagger} - a\right)\left(J_{+} + J_{-}\right) \\ \nonumber
&&+\frac{\omega}{4}\gamma^{2}\left(J_{+} + J_{-}\right)^{2} \\ \nonumber
&&-i\frac{\omega}{2}\sin\left[\gamma\left(a^{\dagger}+ a\right)\right]\left(J_{+} - J_{-}\right)
\end{eqnarray}
To study the critical nature of the system the bosonic and atomic operators have to be modified to allow for macroscopic excitation. This is done  according to the Holstein-Primakoff scheme given in Eq.(\ref{eq.Holstein_Primakoff}) by giving the operators macroscopic displacements as defined in Eq.(\ref{eq.displacement}). In the thermodynamic limit we expand $H_{0}$ in powers of $\sqrt{j}$; this enables the reduction of $H_{m=o}$ to a form :
\begin{equation}
\label{eq.struc_ho_eg}
H_{m=0} = H^{q}_{0}(c,d,c^{\dagger},d^{\dagger}) + \sqrt{j}H^{l}_{0}(c,d,c^{\dagger},d^{\dagger}) + jE_{g}(X,Y),
\end{equation}
where $H^{q}$ contains terms quadratic in $(c,d,c^{\dagger},d^{\dagger})$ and $H^{l}$ contains terms linear in them. Since we only need to analyze the LQE we neglect all terms in the Hamiltonian that contain bosonic operators, namely the linear and quadratic terms mentioned above. Following the method used in \cite{bastidas12}, the structure of the $E_{g}(X,Y)$ surface is used to determine non-equilibrium phase transitions. We obtain the surface as:
\begin{eqnarray}
\label{eq.LQE_sur_kick}
E_{g}(X,Y) =&& \omega X^{2} + \omega(Y^{2}-1)\cos\left(\sqrt{2}X\frac{k\lambda_{1}}{\Omega}\right) \\ \nonumber
&-&\frac{4\lambda_{o}}{\sqrt{2}}XY\sqrt{2-Y^{2}} + \omega\left(\frac{k\lambda_{1}}{2\Omega}\right)^{2}Y^{2}(2-Y^{2}).
\end{eqnarray}

\begin{figure}
\includegraphics[height= 10cm, width=9cm]{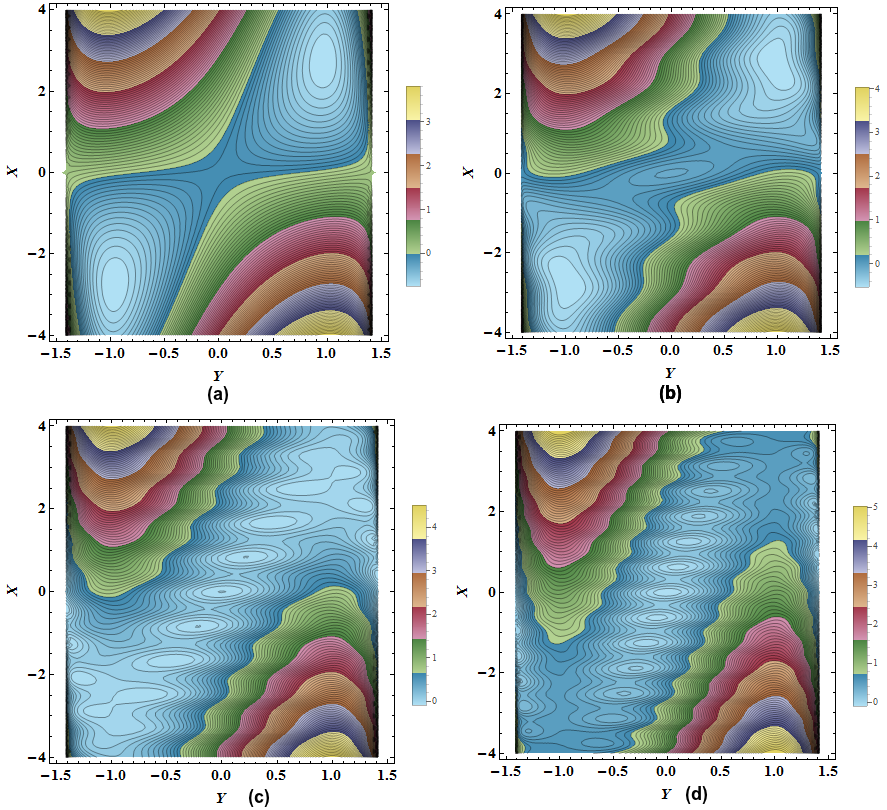}
\caption{(color online)  The quasi energy surface $E_{g}(X,Y)$ is shown for the renormalized kick strength: $k\la_{1} = 0$ (a), $k\la_{1} = 6$ (b), $k\la_{1} = 10.66$ (c), $k\la_{1} = 14$ (d). There is a transition from a scenario with two global minima in (a) representing an SR phase to a scenario with a single global minimum (d) representing the NP. The critical point lies at $k\la_{1} = 10.66$ when a first order phase transition takes place as the values of the order parameter $(X,Y)$ undergo a finite jump to $(0,0)$. }
\label{fig_delta_LQE}
\end{figure}

\begin{figure}
\includegraphics[height= 6cm, width=8cm]{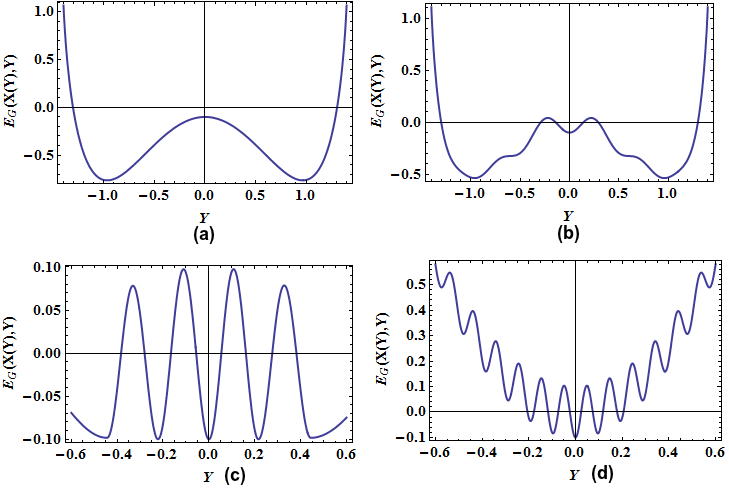}
\caption{(color online) We show above  the cross-sections corresponding to the quasi energy contours in Fig.\ref{fig_delta_LQE}. The first order phase transition and the point of co-existence of the SR and normal phase is clearly seen in panel(c). The behaviour of the LQE observed above is the same as the one reported in \cite{bastidas12}. The difference lies in the fact that unlike the single parameter control of the LQE in the monochromatic case we can now use both the kick strength and the number of kicking cycles to modify the LQE.}
\label{fig_delta_cs}
\end{figure}
In the case of the kicked model we see that Eq.~\eqref{eq.LQE_sur_kick} contains the term $k\lambda_{1}$ where $k$ is the number of $\delta$ kicks and $\lambda_{1}$ is the strength of the kick, thus allowing us to introduce a renormalized kicking strength as $\la = k\lambda_{1}$. This implies that one  can actually modify the surface dynamically by controlling the number of kicks. Setting the parameters of the static system to $\la_{o} = 0.5$, $\omega = .05$ and the driving frequency to $\Omega = 1$ we find that, in the absence of kicks ($k\la_{1}=0$), near the unstable zone (similar to that of the SR phase of the unkicked system) initially $E_{g}$ shows a double global minima at finite $(X,Y)$ as expected (see Fig.\ref{fig_delta_LQE} (a)). On changing the strength $\la$ a local minimum develops at $(X,Y) = (0,0)$ (see Fig.\ref{fig_delta_LQE} (b)), and for a certain value of $\la = 10.66$ we see the presence of three minima, two symmetric ones at finite values of $(X,Y)$ and one at the origin $(X,Y)=(0,0)$, all of the same depth. Recalling that $(X,Y)$ act as the order parameters, one concludes that  the simultaneous presence of minima (in the LQE) of equal depth at both finite and zero values of the order parameter indicate that phases co-exist. Thus, this situation resembles the case of a first order QPT. It is at this point that a so called first order phase transition occurs in the system from a dynamic `Super Radiant' phase to a dynamic `Normal Phase' (see Fig.\ref{fig_delta_LQE} (c)). The local minimum at the origin becomes the global minimum for $\la = 14.0$ (see Fig.\ref{fig_delta_LQE} (d)).\\
Although the behaviour of the LQE observed above is qualitatively the  same as the one reported in \cite{bastidas12}, we would like to emphasize that the difference lies in the fact that unlike the single parameter control of the LQE in the monochromatic case we can now use both the kick strength and the number of kicking cycles to modify the LQE surface.
\begin{figure}
\includegraphics[height= 10cm, width=9cm]{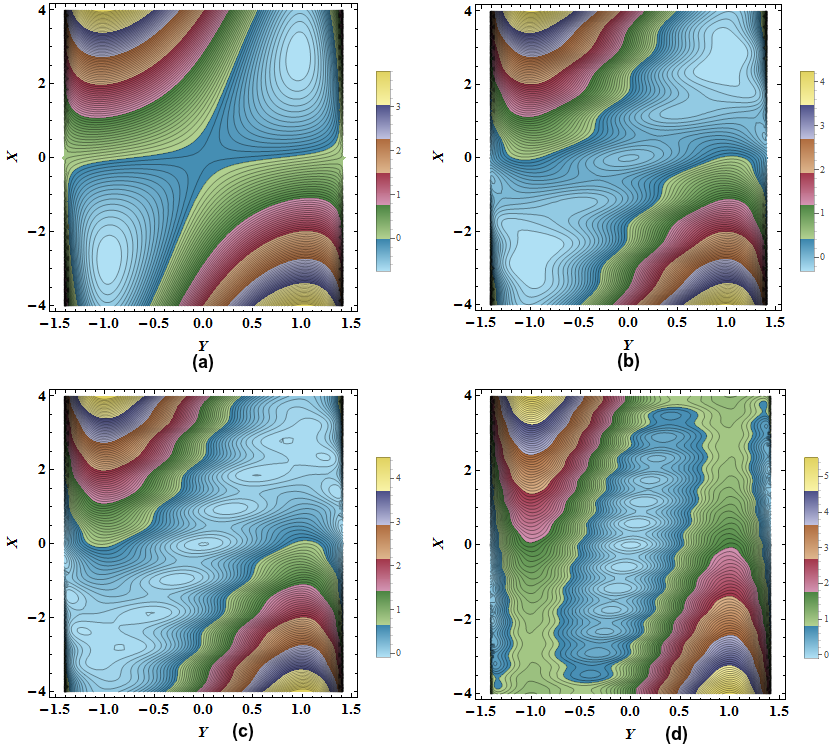}
\caption{(color online) The energy surface plots for the first order phase transition obtained keeping the number of pulses $k = 1$ constant. The panels represent the LQE for various values of the pulse strength (a) $\la_{1} = 0.0$, (b) $\la_{1} = 12.0$, (c) $\la_{1} = 15.02$, (d) $\la_{1} = 24.0$, we can see a distinct crossover of the nature from SR to NP at the critical point represent in (c). }
\label{fig_surface_lam1}
\end{figure}

\begin{figure}
\includegraphics[height= 6cm, width=8cm]{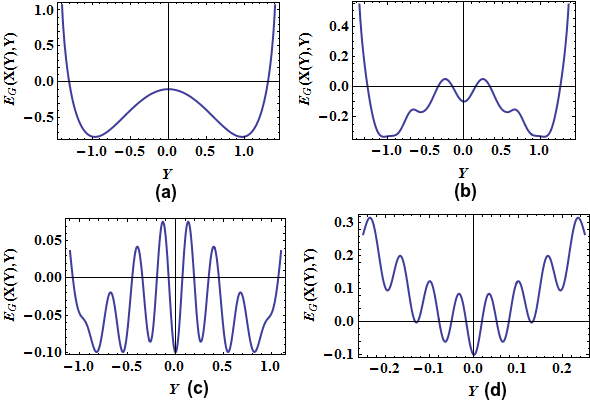}
\caption{(color online)  The energy surface cross-section corresponding to Fig.\ref{fig_surface_lam1}. We can clearly make out the transition and the co-existence point represented in the panel (c).}
\label{fig_lam_cs}
\end{figure}

\begin{figure}
\includegraphics[height= 10cm, width=9cm]{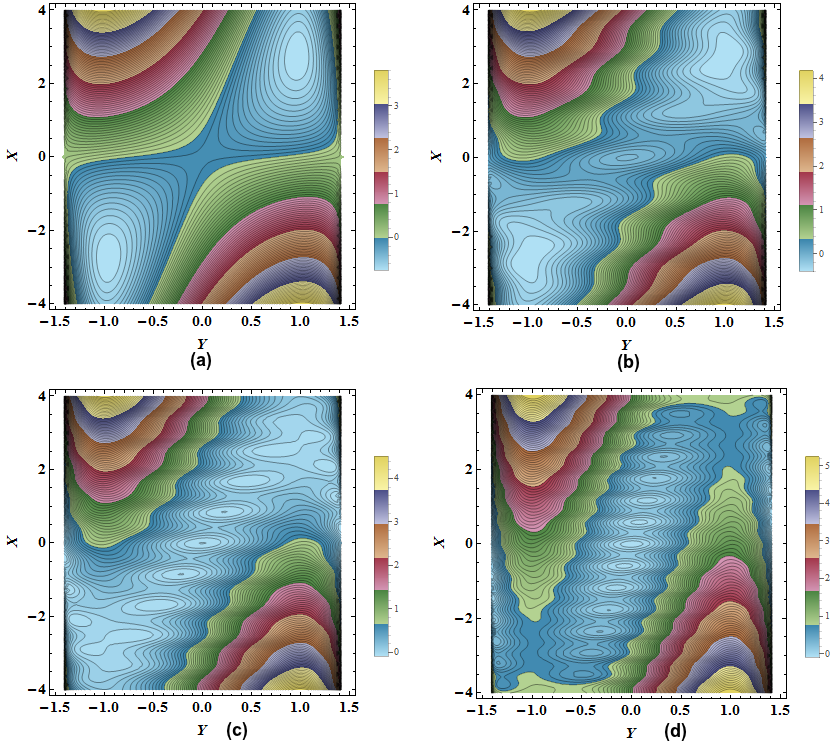}
\caption{(color online)  The energy surface plots for the first order phase transition keeping the pulse strength $\la_{1} = 1$ constant. The panels represent the LQE for various values of the number of pulses $k$ (a) $k = 0$, (b) $k = 12$, (c) $k = 17$, (d) $k = 24$, we can see a distinct crossover of the nature from SR to NP at the critical point represent in (c).}
\label{fig_Kick}
\end{figure}

\begin{figure}
\includegraphics[height= 6cm, width=8cm]{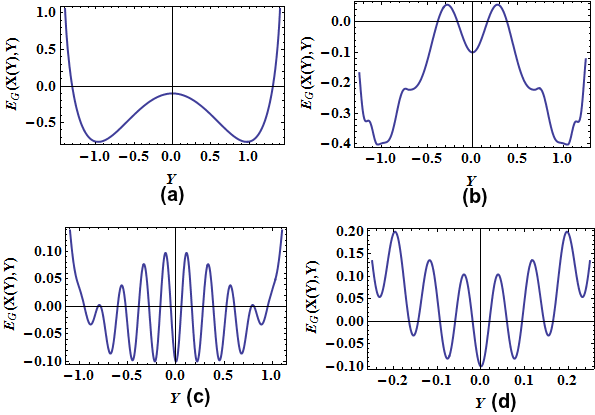}
\caption{(color online)  The energy surface cross-section corresponding to Fig.\ref{fig_Kick}. We can clearly make out the transition and the co-existence point represented in the panel (c).}
\label{fig_kick_cs}
\end{figure}

\section{Bang-Bang Drive}
\label{sec_BB}
The calculation of the LQE for the kicked drive is relatively easy to tackle, however the system is not easily experimentally realizable. Keeping that in mind we use the techniques highlighted in the earlier sections to generalize the LQE structure to that obtained from an asymmetric pulsed drive. In this section we present the calculation and analysis of an asymmetric square wave pulse on the LQE of the DM. The perturbation is a piecewise periodic potential (in time) of the form :
\begin{eqnarray}
\label{eq.V_BB}
V(t) =&& 0; \hspace*{1.75cm} 0<t\le t_{1} \\ \nonumber
V(t) =&& \frac{\lambda_{1}}{\Omega}; \hspace*{1.5cm} t_{1}<t\le T 
\end{eqnarray}
where as before $\lambda_{1}$ is the strength of the pulse and we incorporate the factor $\Omega$ to enable comparison between the LQE obtained in Eq.~(\ref{eq.LQE_sur_kick}) and the present case. $T=2\pi/\Omega$ is the time period of the potential. A stability map similar to that obtained in Eq.~(\ref{eq.map.minus})  can be obtained here as well, and the same criterion for instability namely $\epsilon_{\pm} = m\Omega/2$, is acquired. Like in the previous case we work in the low $\lambda_{o}$ limit near the region $m=0$.\\
To obtain the LQE the same procedure as before is used with a modified structure of the rotation parameter $\gamma$, defined
earlier (see Eq.~(\ref{eq.U_rot})) which in the present form of driving given in Eq.\eqref{eq.V_BB} assumes the form:
 \begin{eqnarray}
\label{eq.gam_BB}
\gamma =&& k\Delta\frac{\lambda_{1}}{\Omega}, \hspace*{3cm} 0<\tau\le t_{1} \\ \nonumber
\gamma =&&  k\Delta\frac{\lambda_{1}}{\Omega} + (t_{1}-\tau)\frac{\lambda_{1}}{\Omega}, \hspace*{1cm} t_{1}< \tau \le T,
\end{eqnarray}
where $\tau = t$ (mod $T$) and $\Delta = T-t_{1}$ and $k$ is the number of pulses delivered to the system. The first of the two $\gamma 's$ in Eq.~(\ref{eq.gam_BB}) simply reproduces our result of the previous section. We hence concentrate on the second expression, writing $\gamma = \gamma_{1}(\Delta , k)+\gamma_{2}(\tau)$ we evaluate the expression derived in Eq.(\ref{eq.flo_ham_modes}),  which necessitates taking the Fourier transforms of $\cos(\gamma_{2}(a+a^{\dagger}))$, $\sin(\gamma_{2}(a+a^{\dagger}))$ and $\gamma_{2}$ to reduce Eq.~(\ref{eq.flo_ham_modes}) to the form $H_{m=0} = \sum_{n=0}^{\infty}h_{m=0}^{n}e^{-in\Omega t}$ (see \cite{bastidas12}). To make the calculations tractable, we use the RWA  limiting ourselves to the $n = 0$ of the Fourier series.\\
To obtain the LQE, as before we displace the bosonic and atomic mode operators of $h^{0}_{0}$ and extract the term without any dependence (quadratic or linear) on the bosonic ($c,c^{\dagger}$) and atomic ($d,d^{\dagger}$) operators. The LQE surface is given by:
\begin{eqnarray}
\label{eq.LQE_bb}
E_{g} =&& \omega X^{2}  \\ \nonumber
&+& \frac{\omega}{4}\left(k\frac{\lambda_{1}}{\Omega}\Delta\right)^{2}Y^{2}(2-Y^{2}) - 2\sqrt{2}\lambda_{o}XY\sqrt{(2-Y^{2})} \\ \nonumber
&+&\frac{\omega(Y^{2} -1)}{T}t_{1}\cos\left(\sqrt{2}kX\frac{\lambda_{1}}{\Omega}\Delta\right)\\ \nonumber
&+& \frac{\Omega\omega(Y^{2} -1)}{\sqrt{2}X\lambda_{1}T}\sin^{2}\left(\frac{\sqrt{2}X\lambda_{1}}{\Omega}\Delta\right)\cos\left(\sqrt{2}kX\frac{\lambda_{1}}{\Omega}\Delta\right)\\ \nonumber
&-&\frac{\sqrt{2}\omega(Y^{2} -1)}{T\lambda_{1}X}\sin\left(\sqrt{2}kX\frac{\lambda_{1}}{\Omega}\Delta\right)\sin^{2}\left(\frac{X\lambda_{1}}{\sqrt{2}\Omega}\Delta\right) \\ \nonumber
&+&\frac{\omega}{T}k\left(\frac{\lambda_{1}}{2\Omega}\right)^{2}\Delta^{3}Y^{2}(2-Y^{2}) \\ \nonumber
&+& \frac{\omega}{4T^{2}}\left(\frac{\lambda_{1}}{2\Omega}\right)^{2}\Delta^{4}Y^{2}(2-Y^{2}) \\ \nonumber
\end{eqnarray}
As can be seen from the structure of Eq.(\ref{eq.LQE_bb}), an asymmetry has been introduced through the term $\Delta = T-t_{1}$ (in time), we also have an asymmetry in the term $k$. Comparing with Eq.~(\ref{eq.LQE_sur_kick}),  we see that unlike the $\delta$-kick
case, for the present driving  one cannot define a renormalized parameter $\lambda = k\lambda_{1}$ since  $\gamma_{2}$ does not involve the parameter  $k$.\\
Let us  now determine the effect of these asymmetries on the surface. It can be clearly seen that in the limit $t_{1} \to T$ and $\lambda_{1}\to \infty$ (keeping $\frac{\lambda_{1}}{\Delta}$ constant), we regenerate the case for the delta kicks as $\frac{1}{\lambda_{1}}\sin\left(\frac{X}{\sqrt{2}\Omega}\lambda_{1}\Delta\right)$, $\Delta^{3}\la_{1}^{2} \to 0$ and $\Delta^{4}\la_{1}^{2} \to 0$. Therefore in the extreme asymmetry limit we get back our LQE for the delta kicks Eq.~(\ref{eq.LQE_sur_kick}) from Eq.~(\ref{eq.LQE_bb}).\\
To further investigate the asymmetries individually we study the effect of each individual component, namely the strength of the perturbation $\lambda_{1}$, the number of pulses $k$ and the asymmetry in time $\Delta$, on the energy surface. Note that in our numerics we retain the same values for $\la_{o}$, $\omega$ and $\Omega$ as used in \ref{sec_rot_frame}.\\
In the first instance, see Figs.\ref{fig_surface_lam1},\ref{fig_lam_cs} , we fix the number of pulses $k =1$ and vary $\lambda_{1}$. We can see that as $\lambda_{1}$ is increased the LQE which was initially showing the characteristics of the SR phase starts developing a minimum at the center of the LQE i.e $(X,Y)=(0,0)$, for $\la_{1} = 12.0$. This central minimum is a characteristic of the NP of the DH. As the strength is increased the central minimum becomes deeper and eventually at a critical value of $\lambda_{1} = 15.02$ the height of the central minimum becomes the same as the heights of the minima which had earlier characterized the SR phase. After this point the system shows a global minimum at $(0,0)$, that is the system enters into the NP and remains there. \\
A similar behaviour takes place as we vary the number of pulses at a fixed strength $\la_{1} =1.0$, see Figs.\ref{fig_Kick}, \ref{fig_kick_cs}. Initially there exists two global minima. A local minimum appears at $(0,0)$ for $k = 12$. The crossover point is located at $k = 17$ at a slightly modified value of the strength $\la_{1} = 0.99$. The strength has to be modified to attain this co-existence region as $k$ can take only integral values, unlike $\la_{1}$. The system is well into the NP for values of $k\ge 19$. Thus we see that we can cause a first order phase transition using both $k$ and $\la_{1}$ individually. However to attain the co-existence region where the heights of the central minimum and the minima at finite $(X,Y)$ are the same we need to fine tune the parameter $\la_{1}$ as $k$ can only take integer values. \\
\begin{figure}
\includegraphics[height= 7cm, width=9cm]{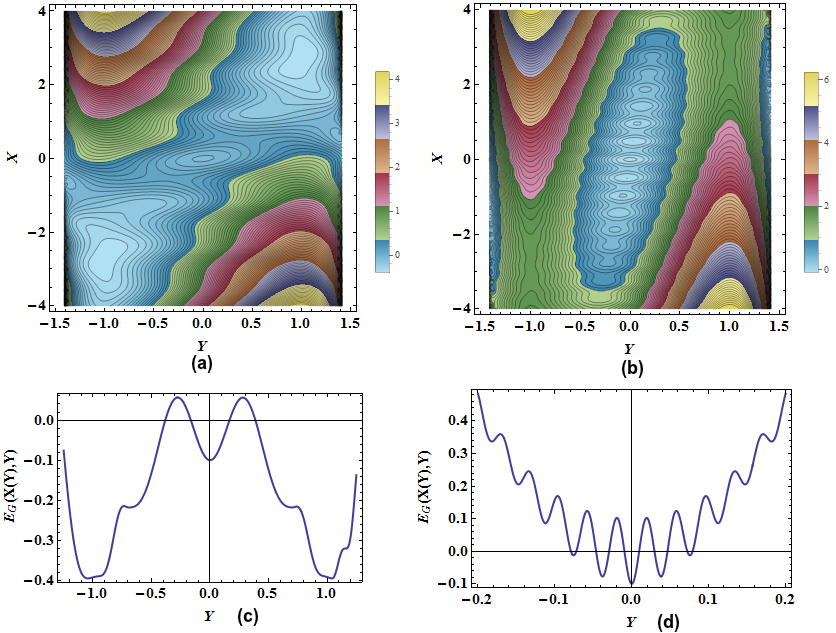}
\caption{(color online) This figure reflects the role of the asymmetry in the pulse width (in time). The parameters are fixed at $\la_{1} = 2$, and $k = 6$. The panel (a),(c) represents the LQE and its cross-section for $\Delta = T/2$, and the panel (b),(d) represents the LQE and its cross-section for $\Delta = T/5$. We can see that the case in which the pulse width is broader there is a phase transition from SR to NP while in the case of the shorter pulse width the system remains in the SR phase.}
\label{fig_time_as}
\end{figure}
As we have introduced a pulse which is asymmetric in time we can investigate the effect of the pulse duration (within a cycle) on the LQE as well, see Fig.\ref{fig_time_as}. For a fixed value of $\la_{1} = 2$ and $k = 6$ we plot the LQE and its cross-section for the case $\Delta = T/2$ in (a) and (c), and $\Delta = T/5$ in (b) and (d). We can clearly see that for a fixed value of $k$ and $\la_{1}$, the case (symmetric) where the pulse duration is longer, a QPT from the SR to the NP is seen while for the pulse of shorter duration the system remains in the initial SR phase. Thus, a change in the pulse-width also has an effect in the possibility of appearance of a first order phase transition.\\
A point to be noted is that if one compares the figure depicting the first order phase transition observed in \cite{bastidas12}, (as shown in Fig.~(3)), to the one we observe in both the kicked and the pulsed DM, is that in the case of \cite{bastidas12} the first metastable states (finite $(X,Y)$) (on either side of the global minimum), have a much smaller depth compared to the global minimum ($(X,Y)=(0,0)$) than in our case where the first metastable states have an $E_{g}$ comparable to the global minimum. This is because in our case the depth of the central minimum is fixed by the magnitude of the cosine term (in Eq.~(\ref{eq.LQE_sur_kick}), and Eq.~(\ref{eq.LQE_bb})) and only after a large number of cycles do the depths of the local minima (at finite $(X,Y)$) become significantly lesser than that of the central minimum. Thus, in the case of the pulsed and kicked DM the probability of the system to remain in a metastable state, is high for a small number of pulse cycles.
\section{Conclusion}
\label{sec_conclusion}
In this paper our main objective was to study a closed quantum system with a pulsed interaction term, in order to investigate the nature of its phases in the presence of such a pulsed drive. We chose the DM as it is an effective model for studying quantum criticality, exhibiting a second order phase transition in equilibrium. We observed that in the thermodynamic limit the DM with a pulsed interaction shows the emergence of a novel phase following from a first order phase transition from the SR phase to the NP. This highlights the point that using external controls like impingement of light, we can dynamically cause a change in the phase, and hence the physical characteristics of a material. In our case the inclusion of a pulsed (or a kicked) term in the interaction term of the DH allows us to `drive' (kick) the system out of a macroscopically excited SR phase to a microscopically excited NP, via a first order phase transition.\\
At this juncture, a comparison with the results of monochromatic perturbation studied earlier would be useful. In the monochromatic case, the phase transition is tuned by varying  a single parameter namely the strength of the drive. In contrary, in the present situation, one can control this transition through both strength and duration of the drive as we have shown in the case of the kicked DM and through strength, duration as well as width (time) of the drive in the case of the pulsed DM. This flexibility enables us to modulate  the phase transitions, and indeed reach the co-existence region through a number of routes. The experimental feasibility of such a phenomenon is increased by the additional dynamical control.

\end{document}